\documentclass[reprint, superscriptaddress, amsmath,amssymb, aps,longbibliography, pra]{revtex4-1}

\usepackage{graphicx}
\usepackage{dcolumn}
\usepackage{bm}
\usepackage{amsmath,amssymb,amstext,amsthm}
\usepackage{braket}
\usepackage{bbold}
\usepackage{bbm}
\usepackage{xcolor}
\usepackage{ gensymb }
\usepackage{soul}
\setstcolor{red}
\usepackage{ mathrsfs }
\usepackage{cancel}
\usepackage{booktabs}

\usepackage{hyperref}
\hypersetup{colorlinks,linkcolor=blue,citecolor=blue,urlcolor=blue}

\newcommand{\pend}{pendell\"osung }

\newcommand{\PCIone}{PCI$_{110}$}
\newcommand{\PCItwo}{PCI$_{350}$}
\newcommand{\PCIthr}{PCI$_{870}$}

\usepackage{etoolbox}

\begin{document}

\preprint{APS/123-QED}

\title{Pendellösung length-scale neutron and X-ray interferometry}

\author{Owen Lailey}
\email{oalailey@uwaterloo.ca}
\affiliation{Institute for Quantum Computing, University of Waterloo, Waterloo, ON, Canada, N2L3G1}
\affiliation{Department of Physics and Astronomy, University of Waterloo, Waterloo, ON, Canada, N2L3G1}

\author{Alexandre Boutot}
\affiliation{Department of Physics and Astronomy, University of Waterloo, Waterloo, ON, Canada, N2L3G1}

\author{David G. Cory}
\affiliation{Institute for Quantum Computing, University of Waterloo, Waterloo, ON, Canada, N2L3G1}
\affiliation{Department of Chemistry, University of Waterloo, Waterloo, ON, Canada, N2L3G1}

\author{Joseph P. Cotter}
\affiliation{Centre for Cold Matter, Blackett Laboratory, Imperial College London, Prince Consort Road, London SW7 2AZ, United Kingdom}

\author{Vishal Dhamgaye}
\affiliation{Diamond Light Source Ltd., Harwell Science \& Innovation Campus, Didcot, Oxfordshire, OX11 0DE, UK}

\author{Tao Hong}
\affiliation{Neutron Scattering Division, Oak Ridge National Laboratory, Oak Ridge, Tennessee 37831, USA}

\author{Michael G. Huber}
\affiliation{National Institute of Standards and Technology, Gaithersburg, Maryland 20899, USA}

\author{Young-June Kim}
\affiliation{Physics Department, University of Toronto, 60 St. George Street, Toronto, Ontario, Canada M5S 1A7}

\author{Winfried Kockelmann}
\affiliation{STFC-Rutherford Appleton Laboratory, ISIS Facility, Harwell OX11 0QX, UK}

\author{Jeremy W. Paster}
\affiliation{National Institute of Standards and Technology, Gaithersburg, Maryland 20899, USA}

\author{Dusan Sarenac}
\affiliation{Department of Physics, University at Buffalo, State University of New York, Buffalo, New York 14260, USA}

\author{Kawal Sawhney}
\affiliation{Diamond Light Source Ltd., Harwell Science \& Innovation Campus, Didcot, Oxfordshire, OX11 0DE, UK}

\author{Naume Shentevski}
\affiliation{Department of Physics, University at Buffalo, State University of New York, Buffalo, New York 14260, USA}

\author{Ivar Taminiau}
\affiliation{Institute for Quantum Computing, University of Waterloo, Waterloo, ON, Canada, N2L3G1}
\affiliation{Neutron Optics Inc., Waterloo, Ontario, Canada M5N 2N1}

\author{Dmitry A. Pushin}
\email{dmitry.pushin@uwaterloo.ca}
\affiliation{Institute for Quantum Computing, University of Waterloo, Waterloo, ON, Canada, N2L3G1}
\affiliation{Department of Physics and Astronomy, University of Waterloo, Waterloo, ON, Canada, N2L3G1}

\date{\today}

\pacs{Valid PACS appear here}

\begin{abstract}
Neutron and X-ray perfect-crystal interferometers (PCIs) are powerful platforms for studies of fundamental physics and phase-contrast imaging. Further enhancing several PCI capabilities requires reducing crystal blade thickness to the micron scale, which minimizes dynamical-diffraction image blur, permits operation in the \pend regime where blade thickness controls beam splitting, and reduces absorption for simultaneous neutron and X-ray operation. However, fabricating multiple crystal blades with identical micrometer-scale thicknesses over centimeter-scale areas remains a major challenge. Here, using a non-etching sub-micron fabrication technique, we demonstrate silicon triple-Laue interferometers with equal-blade-thicknesses of 110~$\mu$m and 350~$\mu$m, operated with both neutrons and X-rays. These devices are the thinnest PCIs realized to date, enabling a factor-of-six reduction in dynamical-diffraction beam spreading for improved phase-contrast imaging, while reaching the single \pend length regime in which crystal thickness provides an experimentally accessible control parameter for engineered quantum-optical beam splitting of plane-wave inputs. These results motivate multi-blade PCI designs utilizing identical half-\pend crystal lamellae that are proposed for neutron spin--orbit and electric dipole moment measurements.

\end{abstract}
\maketitle

\section{Introduction}
Neutron and X-ray perfect crystal interferometers (PCIs) have proven to be powerful tools for studying fundamental physics phenomena and for applications in materials characterization and imaging~\cite{bonse_first, xray_hist, rauch2015neutron, sears}. In particular, X-ray PCI phase-contrast imaging has demonstrated exceptional sensitivity to structural variations, with improvements of up to $\sim10^3$ over absorption-based methods for soft structures~\cite{momose1995demonstration, momose1996phase, momose2003phase2, Momose_2005}, and provides the highest density resolution among existing techniques~\cite{yoneyama2023crystal}. Extensions to neutron phase-contrast imaging similarly offer up to an order-of-magnitude improvement in spatial resolution compared to conventional neutron imaging~\cite{pushin2007reciprocal}. 

The performance of PCI phase-contrast imaging can be further enhanced by reducing crystal thickness. For example, Momose et al. introduced a triple-Laue (LLL) PCI in which the splitter and mirror blades remained thick, while the analyzer blade was thinned to just 40~$\mu$m~\cite{momose1999possibility, momose2003phase}. In this geometry, reducing the crystal thickness suppresses image smearing arising from dynamical diffraction (DD)--induced beam spreading, known as Borrmann fan smearing. 

Beyond their use in phase-contrast imaging, thin crystals provide access to the intrinsic length scales of DD, where the crystal thickness governs the exchange of intensity between transmitted and diffracted beams through the \pend effect~\cite{shull1968observation}. This characteristic length scale, typically on the order of 10--100 micrometers, is defined as:
\begin{equation}
\Delta_H = \frac{1}{C}\frac{\pi V \cos\theta_B}{\lambda|F_H|},
\end{equation}
where $V$ is the unit cell volume, $\theta_B$ is the Bragg angle, $\lambda$ is the wavelength, $C$ is the polarization factor, and $F_H$ is the structure factor. By selecting the crystal thickness according to
\begin{equation}
  t = \Delta_H \times
  \begin{cases}
    (n + 1) & \text{100~\% Transmission} \\
    (n + 1/2) & \text{100~\% Reflection} \\
    (n + 1/4) & \text{50/50 Beam splitter},
    \end{cases}
\label{eq:pend_thick}
\end{equation}
where $n \in \mathbb{N}$, one can realize complete transmission, reflection, or 50/50 beam splitting for incident plane waves, enabling controlled manipulation of neutron and X-ray beam intensities and interferometer visibility~\cite{rauch2015neutron, sears}. Furthermore, thin blades are highly desirable for dual neutron/X-ray interferometry, such as with split-crystal interferometers, where reduced thickness mitigates X-ray absorption and enables high-intensity, gravity-insensitive beams for alignment and test measurements~\cite{lemmel2022neutron}.

Despite these broad advantages in imaging, beam manipulation, and dual neutron/X-ray applications, interferometers composed entirely of identical micron-scale blades remain largely unexplored due to the challenge of fabricating multiple thin lamella with cm$^2$-scale surface areas. Recently, we developed a non-etching, sub-micron fabrication technique for PCIs that achieves uniform phase and near-unity contrast across cm$^2$-scale blade areas~\cite{huber2024achieving}. These advances directly enable the experimental realization of interferometers with blade thicknesses reaching the \pend regime.

\begin{figure}[t]
    \centering
    \includegraphics[width=1\linewidth]{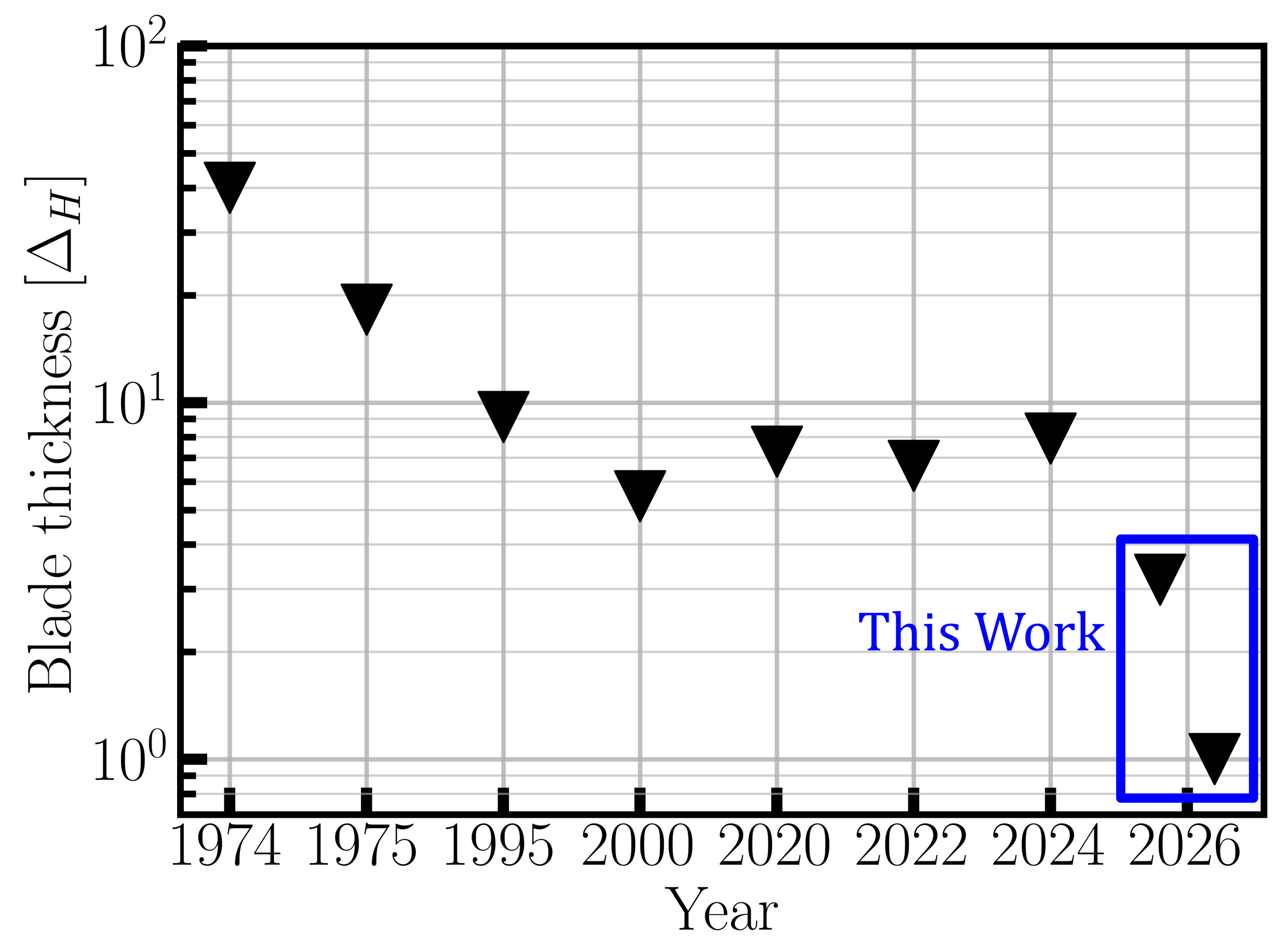}
    \caption{Evolution of equal-blade-thickness triple-Laue (LLL) neutron and X-ray PCIs. Blade thicknesses are expressed in units of the neutron \pend length, $\Delta_H$, for Si (220) at $\lambda=1.3$~\AA. The present work demonstrates the thinnest equal-blade-thickness LLL interferometers reported to date, with $t=110~\mu$m$~=\Delta_H$. References (chronological): first neutron PCI~\cite{rauch1974test}; neutron quantum gravity experiment~\cite{colella1975observation}; X-ray phase-contrast imaging~\cite{momose1995demonstration}; thinnest previously demonstrated equal-blade-thickness LLL PCI~\cite{lin2000angstrom}; split-crystal interferometer~\cite{lemmel2022neutron}; thin-blade strain study~\cite{massa2020x}; non-etching fabrication of LLL interferometers~\cite{huber2024achieving}; and this work.}
    \label{fig:fig1}
\end{figure}

In this work, we demonstrate two high-contrast silicon LLL interferometers with identical blade thicknesses of $t=110~\mu$m and $t=350~\mu$m, operated using both X-rays and neutrons. To the best of the authors' knowledge, these devices are the thinnest equal-blade-thickness triple-blade LLL interferometers realized to date (see Fig.~\ref{fig:fig1}). The $t=110~\mu$m interferometer reaches the single \pend length regime, with $t=\Delta_H$ for the (220) reflection in silicon at a neutron wavelength of $\lambda\sim1.3$~\AA. We benchmark these devices against a previously demonstrated $t=870~\mu$m interferometer~\cite{huber2024achieving}, which achieved 92~\% contrast at the NIST Neutron Interferometry and Optics Facilities~\cite{pushin2015neutron, shahi2016new}. Finally, we use these results to design multi-blade interferometers composed of $N$ identical lamella with $t=\frac{1}{2}\Delta_H$, enabling neutron interferometry platforms proposed for spin-orbit interaction and electric dipole moment measurements~\cite{zeilinger1984symmetry}.

\section{Neutron/X-ray Interferometry}
\label{theory}
\begin{figure*}
    \centering\includegraphics[width=1\linewidth]{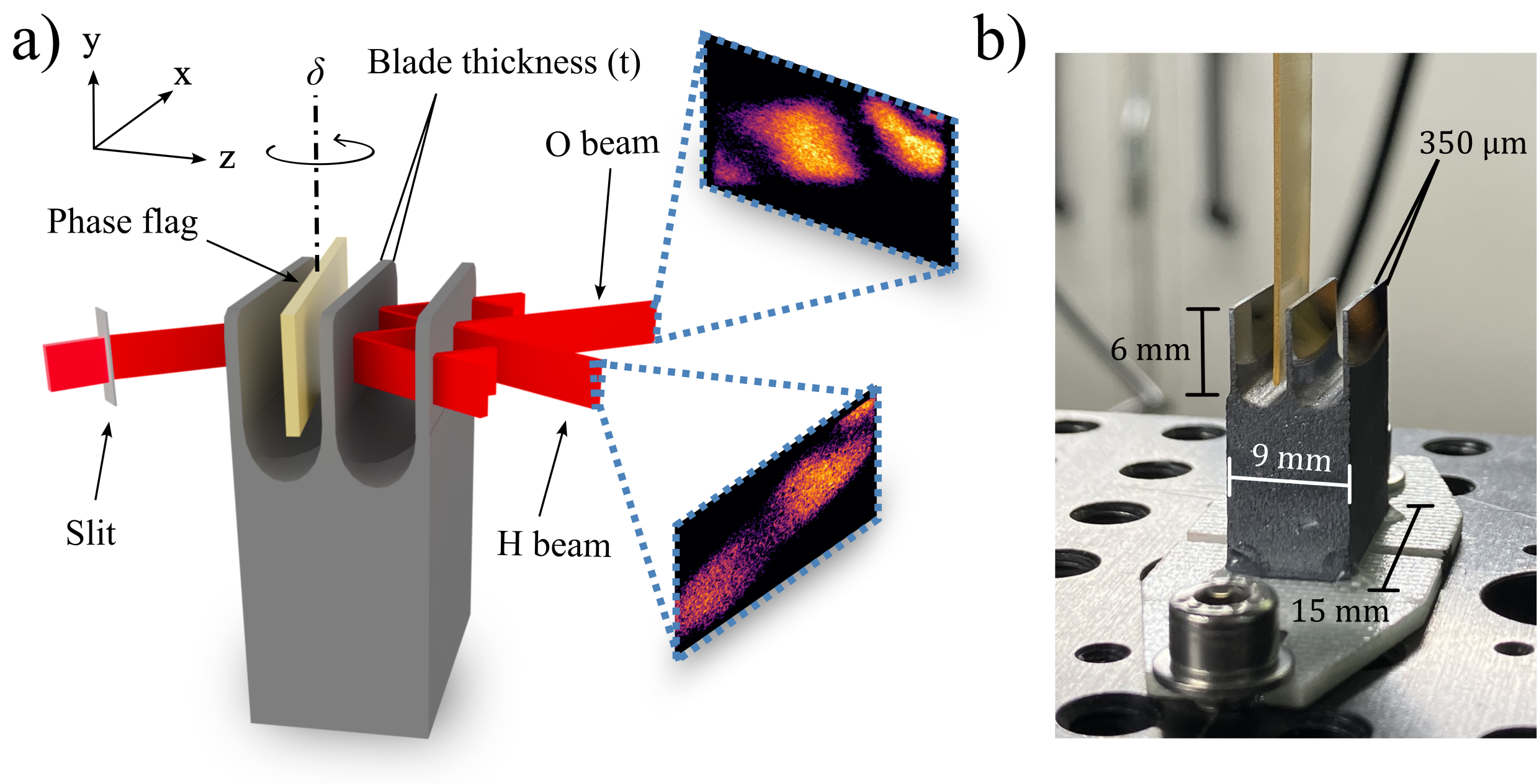}
    \vspace{-20pt}
    \caption{(a) Schematic of the experimental setup for thin-blade neutron and X-ray interferometry. Neutrons/X-rays from an upstream slit are incident on the first interferometer blade of thickness $t$. A phase flag positioned between the first and second blades controls the relative phase between the two~interfering paths. The O- and H-beam intensity profiles are independently recorded by a position-sensitive detector; the inset displays experimental X-ray interference fringes obtained using the $t=110~\mu\text{m}$ (\PCIone) blade interferometer. (b) The $t = 350~\mu\text{m}$ interferometer (\PCItwo) at the Diamond B16 beamline, mounted on an optical breadboard. A phase flag is mounted above the interferometer.}
    \label{fig:fig2}
\end{figure*}

In perfect-crystal interferometry, neutron or X-ray waves are manipulated through Bragg diffraction within a crystalline lattice. Figure~\ref{fig:fig2}a shows a schematic of a LLL interferometer, in which the (220) Bragg planes are oriented perpendicular to the crystal surface. A monochromatic beam, whose size is defined by an upstream slit, enters the first blade of the interferometer and undergoes Bragg diffraction. This process coherently splits the incident wave into two spatially separated components: a forward-diffracted path and a Bragg-reflected path. Subsequent crystal blades redirect these two beams so that some beam paths recombine at a final blade, where interference occurs. This configuration is analogous to a Mach--Zehnder interferometer in optics.

The measured intensities of the output transmitted (O-beam) and reflected (H-beam) depend on the relative phase difference accumulated between the two~paths. This phase difference can arise from materials placed in the beam paths, as well as from external fields or geometric differences in the path lengths. For a uniform sample placed in one~path, the phase shift relative to free-space propagation can be written in general form as
\begin{equation}
\Delta\phi' = k \, \Delta n \, d,
\end{equation}
where $k = 2\pi/\lambda$ is the wavenumber, $\Delta n$ is the difference
in the real part of the refractive index relative to vacuum, and $d$ is the sample thickness. For neutrons, this reduces to $\Delta\phi' = N b_c \lambda d$, where $N b_c$ is the scattering length density. For X-rays, $\Delta n$ is determined by the electron density and atomic form factor.

In practice, a control element, which is known as a phase flag, is placed across both interferometer paths (as shown in Fig.~\ref{fig:fig2}a) and is rotated to vary the effective thickness along each path, thereby modulating the output intensities of the O- and H-beams. The ideal phase flag is as homogeneous as possible to not introduce additional phase differences. The output intensities are integrated over some area and fit to the form
\begin{align}
I(\Delta \phi) &= c_0 + c_1 \cos\left(\Delta \phi + \phi_0\right),
\label{eq:4}
\end{align}
where $c_0$, $c_1$, and $\phi_0$ are fit parameters, with $\phi_0$ being the phase difference between paths for the empty interferometer. The exact expression for $\Delta \phi$ when the phase flag is across both paths of a neutron interferometer and rotated by angle $\delta$ (see Fig.~\ref{fig:fig2}a) is:
\begin{equation}
    \Delta \phi = -2N b_c \lambda d \frac{\sin(\theta_B) \sin(\delta)}{\cos^2(\theta_B) - \sin^2(\delta)},
    \label{bothpaths}
\end{equation}
where $\delta=0$ corresponds to the phase flag parallel to the crystal blades. For small rotations, the differential phase shift becomes approximately linear in $\delta$, enabling a well-controlled phase modulation of the interferometer output.
The contrast, or fringe visibility, provides a measure of interferometer performance and is defined as
\begin{equation}
C = \frac{\text{amplitude}}{\text{mean}} = \frac{c_1}{c_0}.
\label{eq:6}
\end{equation}
Ideally, the O-beam contrast reaches $100~\%$. In practice, however, crystal imperfections, internal strain, surface roughness, geometric misalignments, absorption, and environmental noise reduce the achievable contrast.

With a 2D position-sensitive detector (as shown in Fig.~\ref{fig:fig2}a), multiple interference fringes can be observed simultaneously across the detector. These spatial fringes arise from a phase gradient across the interfering beams, which may result from several factors, including a wedged phase flag, internal strain/stress, or temperature gradients. In such cases, the fringe contrast can be extracted from a single-shot measurement, without requiring a full phase-flag rotation scan. Additionally, to demonstrate controlled manipulation of the interference pattern, rotating the phase flag to produce a relative phase shift of approximately $\Delta\phi \approx \pi$ will shift the pattern by half a fringe period. A normalized intensity difference between the two measurements can then be defined as
\begin{equation}
C_{\pi} = \frac{I(\pi) - I(0)}{I(\pi) + I(0)},
\label{intens_ratio}
\end{equation}
where $I(\Delta\phi)$ is given in Eq.~\ref{eq:4}. This quantity provides a practical lower bound estimate to the contrast definition in Eq.~\ref{eq:6} ($C_\pi=C\cos(\phi_0)$), requiring only two measurements rather than a full phase-flag scan. This approach is particularly advantageous when phase-flag rotation is restricted or when poor environmental conditions, combined with low count rates, cause the interference fringes to wash out during an extended scan. Under such conditions, the one or two-measurement approach can provide a more robust estimate of the intrinsic interferometer contrast.

This limited-measurement approach is conceptually related to quantum Hamiltonian learning, in which properties of an underlying quantum system are inferred from a relatively small number of appropriately chosen measurements~\cite{granade2012robust,ferrie2013best,wiebe2014quantum}. Similarly, the two-measurement procedure here extracts the interferometer contrast without requiring a complete phase-flag scan, reducing the number of measurements needed to characterize the system.

\section{Materials and Methods}
Three LLL-type perfect silicon crystal interferometers were measured in this work. The largest device, ``Talos,'' with a blade thickness of \(t = 870~\mu\text{m}\) and operating in the (111) reflection, has been previously characterized and was used here as a reference for the smaller interferometers under identical experimental conditions. As described in Ref.~\cite{huber2024achieving}, iterative submicrometer, non-etching fabrication techniques were used to improve interferometer contrast. Under controlled temperature and vibration conditions at the NIST Neutron Interferometry and Optics Facilities~\cite{pushin2015neutron, shahi2016new}, Talos achieved contrast values up to \(92~\%\) with uniform phase across the blade surface.

The smaller interferometers are characterized here for the first time. They were fabricated using the same methods as Talos and have blade thicknesses of \(t = 110~\mu\text{m}\) and \(t = 350~\mu\text{m}\), operating in the (220) reflection. To our knowledge, these are the thinnest neutron and X-ray interferometers fabricated to date with all three lamellae having equal thickness. Previously, a single \(40~\mu\text{m}\)-thick blade was fabricated for use in a neutron interferometer~\cite{momose1999possibility, momose2003phase}. Each interferometer has base dimensions of \(15~\text{mm} \times 9~\text{mm}\), a base height of approximately \(14~\text{mm}\), and blades extending an additional \(6~\text{mm}\). The blade separation between center lines is \(4~\text{mm}\). An image of the \(t = 350~\mu\text{m}\) interferometer is shown in Fig.~\ref{fig:fig2}b.

Throughout the text the interferometers are referred to as $\text{PCI}_t$, where $t = 110~(\mu$m), $350~(\mu$m), and $870~(\mu$m) for the 3 interferometers.

All interferometers were tested in multiple configurations. All measurements were conducted without active vibration or temperature isolation. This work therefore also represents the first operation of \PCIthr~outside NIST without environmental control. As a result, the baseline contrast values reported here for \PCIone~and \PCItwo~are conservative; higher contrast is expected under controlled conditions.

The experiments were performed in the following four configurations:

{
\setcounter{subsection}{0}
\renewcommand{\thesubsection}{\roman{subsection}}
\subsection{X-Ray Diffractometer}
The first configuration used Cu K$\alpha$ x-rays with wavelength $\lambda = 1.54$~\AA~from a high-resolution Rigaku SmartLab X-ray diffractometer. H-beam interference fringes were measured using a 2D position-sensitive detector with a wedged phase flag inserted between the first and second blades across both beam paths. The phase flag was made of PMMA with a wedge angle of 2~\degree. Rotation of the phase flag was performed manually, which limited the ability to achieve an exact $\Delta\phi = \pi$ condition required for Eq.~\ref{intens_ratio}. Measurements were conducted for \PCIone~and \PCItwo. The incident beam size was $2~\text{mm} \times 1~\text{mm}$, and images were acquired with an exposure time of 60~s.

\subsection{X-Ray Synchrotron}
The second configuration was carried out at the Diamond Light Source Test beamline B16~\cite{sawhney2010test} in monochromatic mode at 8~keV and 15~keV. Both O-beam and H-beam interference profiles were measured using a flat phase flag inserted between the first and second blades across both beam paths. A Photonic Science MiniFDI detector was used, providing an 11~mm diameter field of view with $6.5~\mu\text{m}$ pixel size. The phase flag, made of Ultem (PEI)~\cite{nist_disclaimer} with a thickness of 1~mm, was rotated using an automated stage over a range of approximately 12~\degree. Measurements were performed for all three interferometers with exposure times ranging from 0.1~s to 2~s. In this preliminary study, insufficient environmental isolation resulted in phase drift during the measurement. To reduce the sensitivity to this drift, the interferometric contrast was determined using two complementary approaches: phase-flag rotation (Eq.~\ref{eq:4}) and single-shot fringe analysis (Eq.~\ref{intens_ratio}).

\subsection{Neutron Imaging}
The third configuration was performed at the ISIS Neutron and Muon Source imaging beamline IMAT~\cite{kockelmann2013imat}. Measurements were conducted in time-of-flight (TOF) mode with a polychromatic beam, allowing variable wavelength selection by rotation of the interferometer (see Appendix Fig.~\ref{fig:fig7}). A Berkeley MCP (Timepix 2) detector was used, with a $28$~mm$~\times~28~\text{mm}$ field of view and $55~\mu\text{m}$ pixel size. O-beam intensity profiles were measured for all three interferometers.

For \PCIone, a $1$~mm$~\times~14~\text{mm}$ upstream slit and a 52~minute acquisition time were used at a Bragg angle of $40^\circ$. For \PCItwo, a $1$~mm$~\times~19~\text{mm}$ slit and a 1~h acquisition time were used at $55^\circ$. For \PCIthr, a $1$~mm$~\times~10~\text{mm}$ slit and a 2~h acquisition time were used at $26^\circ$. A full contrast measurement was performed only for \PCIthr~using $\lambda = 2.78$~\AA~and a fused silica phase flag of thickness 1.675~mm. This measurement was conducted over 11~h with manual phase flag rotations, while temperature and vibration were monitored, but not controlled, adjacent to the interferometer (see Appendix Fig.~\ref{fig:fig8}).

\begin{figure*}
    \centering\includegraphics[width=1\linewidth]{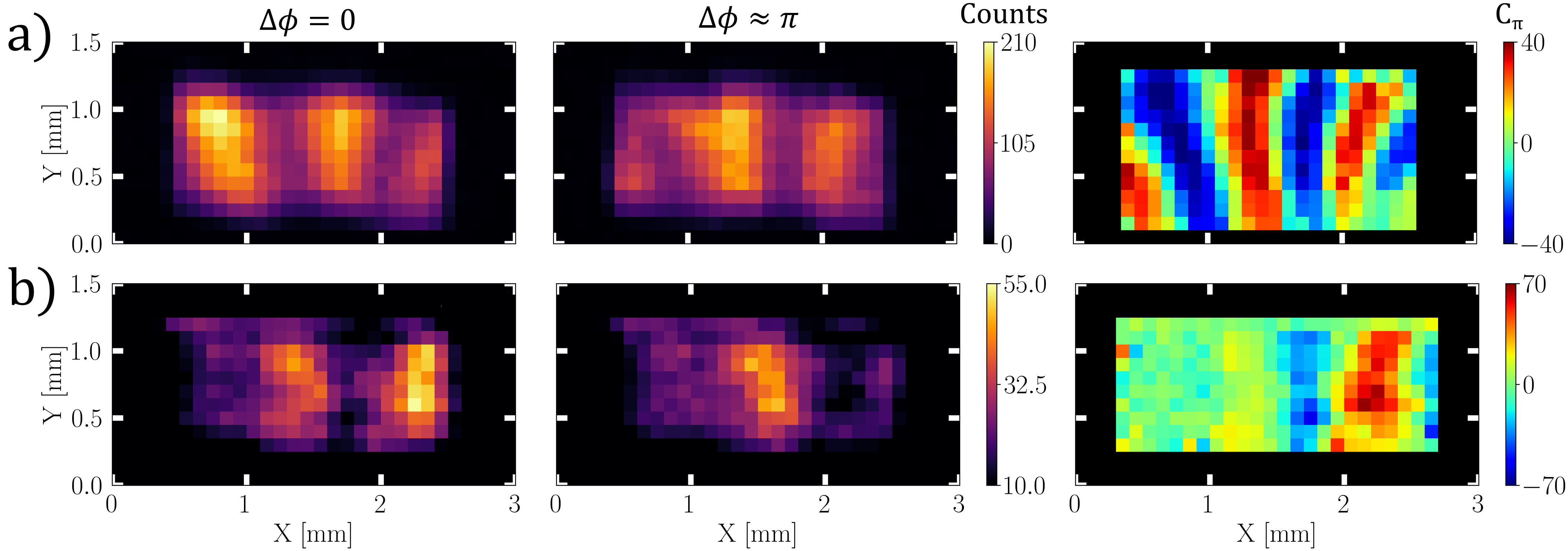}
    \caption{(a) X-ray interference fringes obtained using \PCIone~at the X-ray diffractometer using the wedged phase flag. The phase flag was manually rotated, demonstrating $\Delta\phi \approx \pi$ between the two~images. We compute the normalized intensity difference maps $C_\pi$ (Eq.~\ref{intens_ratio}), indicating a contrast of approximately $C_\pi = 38~\% \pm 6~\%$. (b) X-ray interference fringes obtained using \PCItwo~at the X-ray diffractometer using the wedged phase flag. The normalized intensity difference maps indicates a contrast of approximately $C_\pi = 67~\% \pm 10~\%$. In both cases, the $C_\pi$ maps are calculated from only two measurements separated by $\Delta\phi \approx \pi$ and therefore should not be interpreted as conventional contrast maps obtained from scanning across the interferometer blade surfaces.}
    \label{fig:fig3}
\end{figure*}

\subsection{Neutron Triple-Axis Spectrometer}
The fourth configuration was performed at the ORNL HFIR triple-axis spectrometer CTAX. O-beam contrast was measured using an integrating $^3$He detector and a flat sapphire phase flag. Due to time constraints, a full contrast measurement was performed only for \PCItwo. The phase flag had a thickness of $450~\mu\text{m}$ and was rotated over $5^\circ$ in $0.25^\circ$ steps, with 60~s acquisition per step. The neutron wavelength was 3.0~\AA, the Bragg angle was $51.62^\circ$, and the beam size was $1$~mm$~\times~4~\text{mm}$.
}

\section{Results}

\subsection{X-ray Measurements}
Shown in Fig.~\ref{fig:fig3} are the measured 2D H-beam interference profiles for (a) \PCIone~and (b) \PCItwo~obtained with the laboratory X-ray diffractometer. A wedged phase flag was inserted across both interferometer paths, introducing a transverse phase gradient across each $2~\text{mm} \times 1~\text{mm}$ beam. The phase flag was rotated to produce an approximate $\Delta\phi \approx \pi$ shift of the fringes between the two~measurements, as shown in the first and second columns of Fig.~\ref{fig:fig3}. 

Approximately two fringe periods are observed across the 2~mm beam width in all images. Taking a horizontal slice at Y = 1.0 mm in the \PCIone~$\Delta\phi = 0$ image and fitting to a cosine yields a fringe period of $1.02 \pm 0.04$~mm. This is in close agreement with the expected fringe period of approximately 1.08~mm for the implemented $2^\circ$ PMMA wedge. The interference fringes exhibit both angled and curved features, which may arise in part from thickness variations in the wedged phase flag, which was not fabricated as a precision optic. More broadly, such distortions are commonly observed in X-ray and neutron interferometers and can result from residual lattice strain, temperature gradients, angular misalignment, phase-flag misalignment, or mounting-induced stress. Using the two~measurements with $\Delta\phi \approx 0$ and $\Delta\phi \approx \pi$, the normalized intensity ratio (Eq.~\ref{intens_ratio}) was used to extract the fringe contrast, as shown in the third column of Fig.~\ref{fig:fig3}. The maximum contrast values are $C_\pi = 38~\% \pm 6~\%$ for \PCIone~and $C_\pi = 67~\% \pm 10~\%$ for \PCItwo, indicating successful interferometer functionality. 

\begin{figure*}
    \centering\includegraphics[width=1\linewidth]{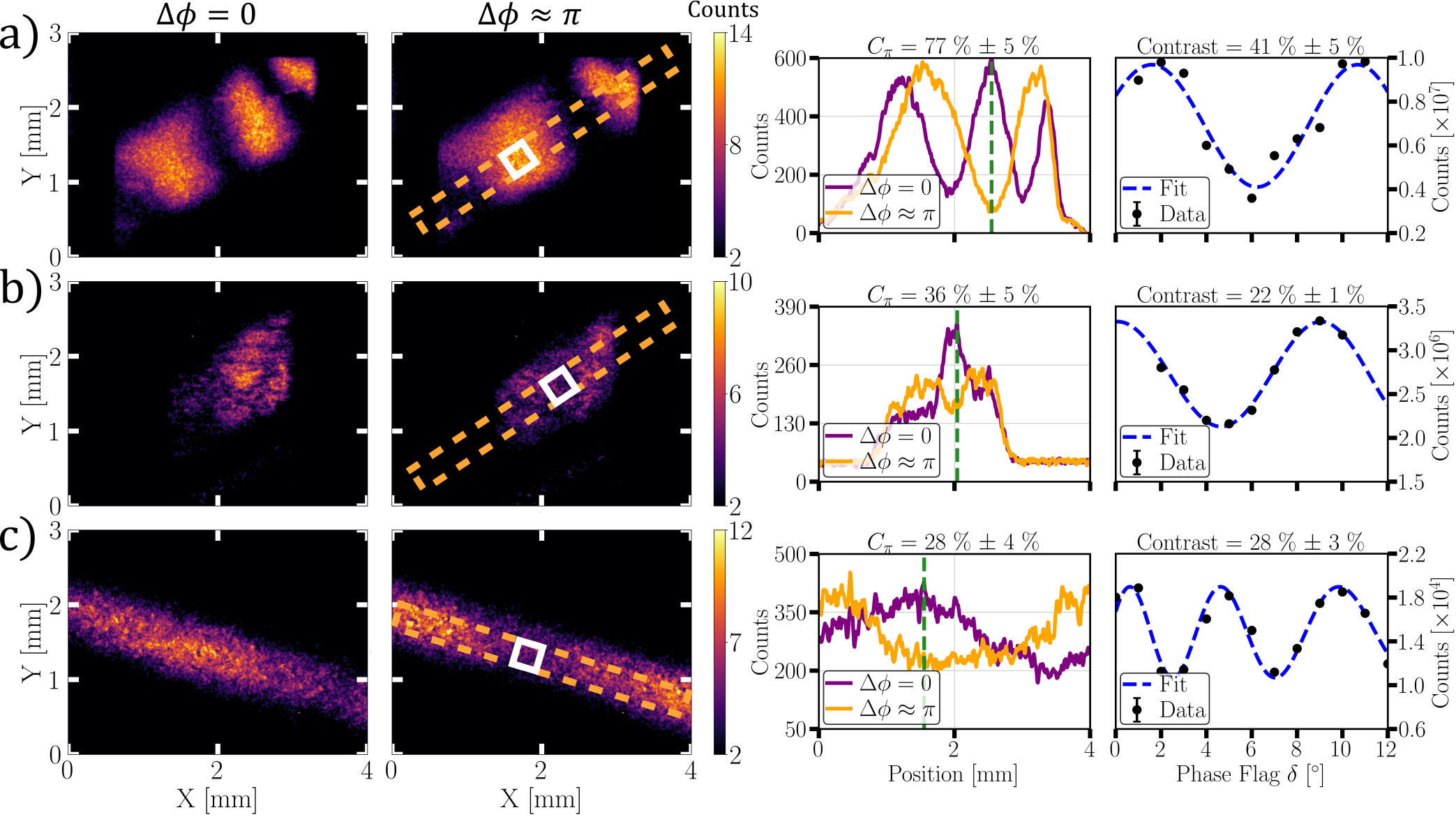}
    \caption{(a) O-Beam X-ray interference fringes obtained with \PCIone. Consecutive images were acquired with 0.1~s exposure at 8~keV. The apparent $\sim\pi$ phase shift between images arises from phase drift due to limited environmental stability. The transverse intensity slice (third column), obtained by integrating over the 50-pixel-wide orange rectangle, yields a contrast of $77~\% \pm 5~\%$ at the position indicated by the dashed green line. Due to the fringe curvature, locally higher contrast ($\gtrsim 80~\%$) can be observed over narrower integration regions. A full phase scan (fourth column) was performed at 15~keV using a flat phase flag, yielding a reduced contrast of $41~\% \pm 5~\%$. The scan integrates counts over the $50 \times 50$ pixel white square. (b) O-Beam interference fringes for \PCItwo. Consecutive images were acquired with 2~s exposure at 8~keV. As in (a), the apparent phase shift between frames is due to phase drift. The full $2\pi$ phase scan (fourth column), performed at 15~keV, shows reduced contrast under the same conditions. (c) Interference fringes for \PCIthr. Images were acquired with 0.5~s exposure at 15~keV. In this case, the phase flag was inserted into a single path, resulting in nonuniform fringe spacing during the phase scan. All intensity profiles in (a)--(c) are background-subtracted and smoothed with a Gaussian kernel ($\sigma = 1$ pixel $= 6.5~\mu\text{m}$). Some error bars, representing $1\sigma$ Poisson counting statistics, lie within the markers.}
    \label{fig:fig4}
\end{figure*}

All three interferometers were further characterized with X-rays at the Diamond Light Source. Fringe contrast was analyzed using both methods described in Sec.~\ref{theory}. Figure~\ref{fig:fig4}(a)--(c) shows O-beam interference fringes for \PCIone, \PCItwo, and \PCIthr, respectively. In these measurements, apparent $\sim\pi$ phase shifts between consecutive images arise from rapid phase drift in the interferometers due to the absence of environmental isolation. The interferometers were exposed to ambient vibrations, temperature fluctuations, and uncontrolled air flow within the hutch, leading to $\pi$ phase variations on timescales as short as 0.1~s.

To estimate the fringe contrast under these conditions, consecutive images with an approximate $\pi$ phase difference were selected. Transverse line profiles (third column of Fig.~\ref{fig:fig4}) were obtained by integrating over the 50-pixel-wide short axis of the dotted, orange rectangles overlaid on the fringes. The contrast was then calculated using Eq.~\ref{intens_ratio}, yielding $C_\pi = 77~\% \pm 5~\%$, $36~\% \pm 5~\%$, and $28~\% \pm 4~\%$ for \PCIone, \PCItwo, and \PCIthr, respectively. These values are summarized in Table.~\ref{tab:contrast}.

\begin{table}[t]
    \centering
    \caption{Measured interferometric contrast for the three interferometers in the various configurations demonstrated in Figs.~\ref{fig:fig3},~\ref{fig:fig4},~\ref{fig:fig5}. Computed contrast values are given by Eqs.~\ref{eq:6}, ~\ref{intens_ratio}.}
    \label{tab:contrast}
    \begin{tabular*}{\columnwidth}{@{\extracolsep{\fill}}l||ccc@{}}
        \toprule
        Measurement & \PCIone & \PCItwo & \PCIthr \\
        \hline \hline
        \midrule
        X-rays, Fig.~\ref{fig:fig3}: $C_\pi$ [\%] & $38 \pm 6$ & $67 \pm 10$ & N/A \\
        X-rays, Fig.~\ref{fig:fig4}: $C_\pi$ [\%] & $77 \pm 5$ & $36 \pm 5$ & $28 \pm 4$ \\
        X-rays, Fig.~\ref{fig:fig4}: $C$ [\%] & $41 \pm 5$ & $22 \pm 1$ & $28 \pm 3$ \\
        Neutrons, Fig.~\ref{fig:fig5}: $C$ [\%] & N/A & $8 \pm 3$ & $58 \pm 9$ \\
        \bottomrule
    \end{tabular*}
\end{table}

A conventional contrast scan was also performed by rotating a flat phase flag. Due to the same environmental instability, reduced contrast was observed. The integrated intensity was evaluated over the 50-pixel-wide region indicated by the white square and fit using Eq.~\ref{eq:4} and Eq.~\ref{bothpaths}. The resulting contrast values are $41~\% \pm 5~\%$ for \PCIone~and $22~\% \pm 1~\%$ for \PCItwo. The fitted phase-flag thicknesses are $917~\mu$m $\pm~80~\mu\text{m}$ and $942~\mu$m $\pm~27~\mu\text{m}$, respectively, consistent with the independently measured thickness of 1~mm.

For \PCIthr~(Fig.~\ref{fig:fig4}(c)), the phase flag did not span both interferometer paths and was instead inserted into a single arm. In addition, the rotation was not centered about $\delta = 0$. As a result, the observed oscillations have a higher and varying spatial frequency, consistent with the single-arm phase flag dependence $\Delta\phi \propto 1/\cos(\theta_B - \delta)$. The extracted contrast in this configuration is $28~\% \pm 3~\%$.

\subsection{Neutron Measurements}
\begin{figure}
    \centering\includegraphics[width=1\linewidth]{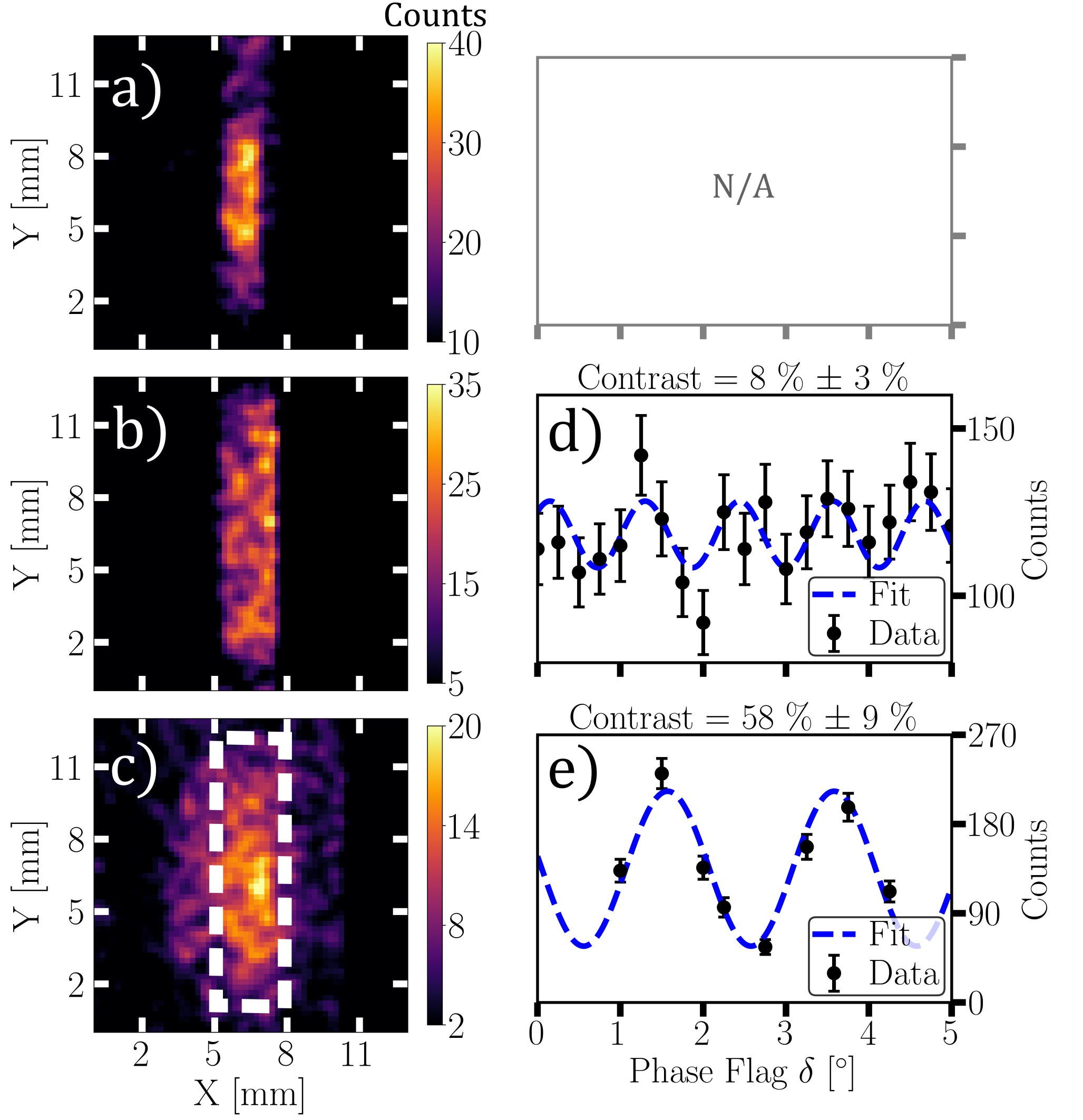}
    \caption{(a)--(c) Neutron O-Beam images for \PCIone, \PCItwo, and \PCIthr, integrated across a wavelength range of 0.09~\AA, 0.06~\AA, and 0.03~\AA~respectively. Images are binned in 4x4 pixel regions ($1\text{ pixel} = 55~\mu\text{m}$), background-subtracted and filtered using a Gaussian kernel with $\sigma = 220~\mu\text{m}$. A neutron contrast measurement was not performed for \PCIone. (d) Neutron contrast measurement for \PCItwo~obtained at CTAX at ORNL. Uncertainties represent $1\sigma$ Poisson counting statistics. (e) Neutron contrast measurement for \PCIthr~obtained at IMAT using the neutron imaging setup. The dotted white rectangle in (c) indicates the integration region used for computing contrast.}
    \label{fig:fig5}
\end{figure}

In the neutron imaging configuration, the O-beam intensity profile was measured for \PCIone, \PCItwo, and \PCIthr, as shown in Fig.~\ref{fig:fig5}(a)--(c), respectively. Long acquisition times were required due to the low neutron count rate of $\sim 0.1$~Hz within a $0.003~\text{\AA}$ wavelength bin at the Bragg peak (see Appendix Fig.~\ref{fig:fig7}). All images were acquired using a 1~mm slit width in the horizontal ($x$) direction.

Across Fig.~\ref{fig:fig5}(a)--(c), a progressive broadening of the O-beam is observed, consistent with dynamical diffraction (DD) Borrmann fan spreading within the crystal blades. Each blade contributes a transverse spread of $\Delta l = 2t \sin\theta_B$, which accumulates across the three blades. This yields expected total beam widths of $1.4$~mm, $2.7$~mm, and $3.3$~mm for \PCIone, \PCItwo, and \PCIthr, respectively. Experimentally, horizontal line profiles were extracted and fit with a smoothed top-hat function, giving widths of $1.5~\text{mm} \pm 0.1~\text{mm}$, $2.6~\text{mm} \pm 0.1~\text{mm}$, and $3.2~\text{mm} \pm 0.3~\text{mm}$, in good agreement with these estimates. The images were integrated over wavelength bandwidths of $\sim 0.09~\text{\AA}$, $\sim 0.06~\text{\AA}$, and $\sim 0.03~\text{\AA}$ for (a)--(c), respectively, to increase counting statistics; this integration can introduce additional broadening due to the corresponding spread in Bragg angle. These results directly demonstrate the importance of thinner blade interferometers (\PCIone~$\sim8\times$ smaller than \PCIthr) for reducing the image blur $\Delta l = 2t\sin\theta$ associated with DD beam spreading.

As discussed, spatial variations in the O-beam intensity may arise from phase gradients (e.g., from a wedged optic), environmental perturbations, or crystal strain. In the absence of resolved interference fringes in Fig.~\ref{fig:fig5}, such effects may manifest as intensity variations exceeding counting statistics. To assess this, the mean intensity $\mu$ (counts) and standard deviation $\sigma$ were computed within the O-beam region, yielding: (a) $\mu = 28$, $\sigma = 5.7$; (b) $\mu = 19.3$, $\sigma = 4.7$; (c) $\mu = 10.6$, $\sigma = 3.4$. In all cases, $\sigma \approx \sqrt{\mu}$, consistent with Poisson statistics, indicating no measurable excess fluctuations above counting noise within the sensitivity of these measurements.

Contrast measurements for \PCItwo\ and \PCIthr\ are shown in Fig.~\ref{fig:fig5}(d,e) and summarized in Table~\ref{tab:contrast}. Due to time constraints, a full contrast measurement was not performed for \PCIone. The \PCItwo\ measurement was carried out at a triple-axis spectrometer using an integrating $^3$He detector. The data were fit using Eq.~\ref{eq:4} and Eq.~\ref{bothpaths}, with the sapphire phase flag thickness as a free parameter. This yields a contrast of $8~\% \pm 3~\%$ and a fitted thickness $d = 446~\mu\text{m} \pm 15~\mu\text{m}$, in agreement with the measured value of $450~\mu\text{m}$.

The contrast measurement for \PCIthr\ (Fig.~\ref{fig:fig5}(e)) was performed using the 2D neutron imaging setup. Fitting with Eq.~\ref{eq:4} and Eq.~\ref{bothpaths}, using the fused silica phase flag thickness as a free parameter, gives a contrast of $58~\% \pm 9~\%$ and $d = 1.665~\text{mm} \pm 0.063~\text{mm}$, consistent with the measured thickness of 1.675~mm. This contrast is lower than the $92~\%$ reported in Ref.~\cite{huber2024achieving}, primarily due to the lack of vibration and temperature isolation. The long acquisition times increase sensitivity to phase drift, reducing the measured contrast. In addition, the phase flag rotation was performed manually, requiring periodic entry into the hutch and introducing disturbances. Temperature and vibration were monitored adjacent to the interferometer, showing spikes of $\sim 0.1^\circ$C and $\sim 0.2$~m/s$^2$ during phase flag rotations (see Appendix Fig.~\ref{fig:fig8}).

\section{Discussion}
The results presented here demonstrate that high-quality neutron and X-ray interference can be sustained in perfect crystal interferometers with blade thicknesses at the \pend length scale. In this regime, the role of DD becomes increasingly explicit: crystal thickness not only influences beam broadening but also determines the rotation of the beam-splitting transformation, thereby controlling interferometer visibility. 

From an imaging perspective, reducing the blade thickness suppresses Borrmann fan spreading, thereby improving spatial resolution and preserving transverse phase information. This is particularly relevant for neutron phase-contrast imaging, where DD-induced blurring is a primary limitation in conventional, millimeter-scale interferometers~\cite{becker2001neutron}. The $t = 110~\mu$m interferometer represents an approximately $8\times$ reduction in Borrmann fan broadening compared to the $t = 870~\mu$m interferometer, and an approximate $6\times$ reduction for the previously thinnest LLL interferometer with $t = 600~\mu$m~\cite{lin2000angstrom}. This substantial decrease in DD-induced beam spreading directly improves the preservation of transverse phase information and highlights the advantage of operating with pendell\"osung-scale blade thicknesses.

At the same time, reaching $t = \Delta_H$ opens access to regimes where beam splitting can be engineered through crystal thickness, transforming the blade into a controllable quantum-optical element. The evolution of the transmitted and diffracted components acts as a thickness-dependent ($2\times2$) unitary transformation, expanding the role of the \pend length from merely describing intensity oscillations to setting the rotation scale between the two modes. This direct connection to quantum-optical interferometry means that crystal thickness establishes the path amplitudes; an optimal thickness balances these amplitudes to maximize fringe visibility, enabling engineered splitting ratios in multi-blade perfect-crystal interferometers. However, while ideal DD predicts $100~\%$ reflection or transmission at specific fractions of $\Delta_H$, these conditions are strictly valid only for monochromatic plane waves~\cite{sears, rauch2015neutron}. In realistic experimental settings, finite wavelength bandwidth and angular divergence lead to a distribution of \pend phases, reducing the achievable contrast and smoothing the ideal response. Nevertheless, the interferometers measured in this work indicate that this regime remains experimentally accessible and useful for designing interferometers with tailored splitting ratios.

\begin{figure}
    \centering
    \includegraphics[width=1\linewidth]{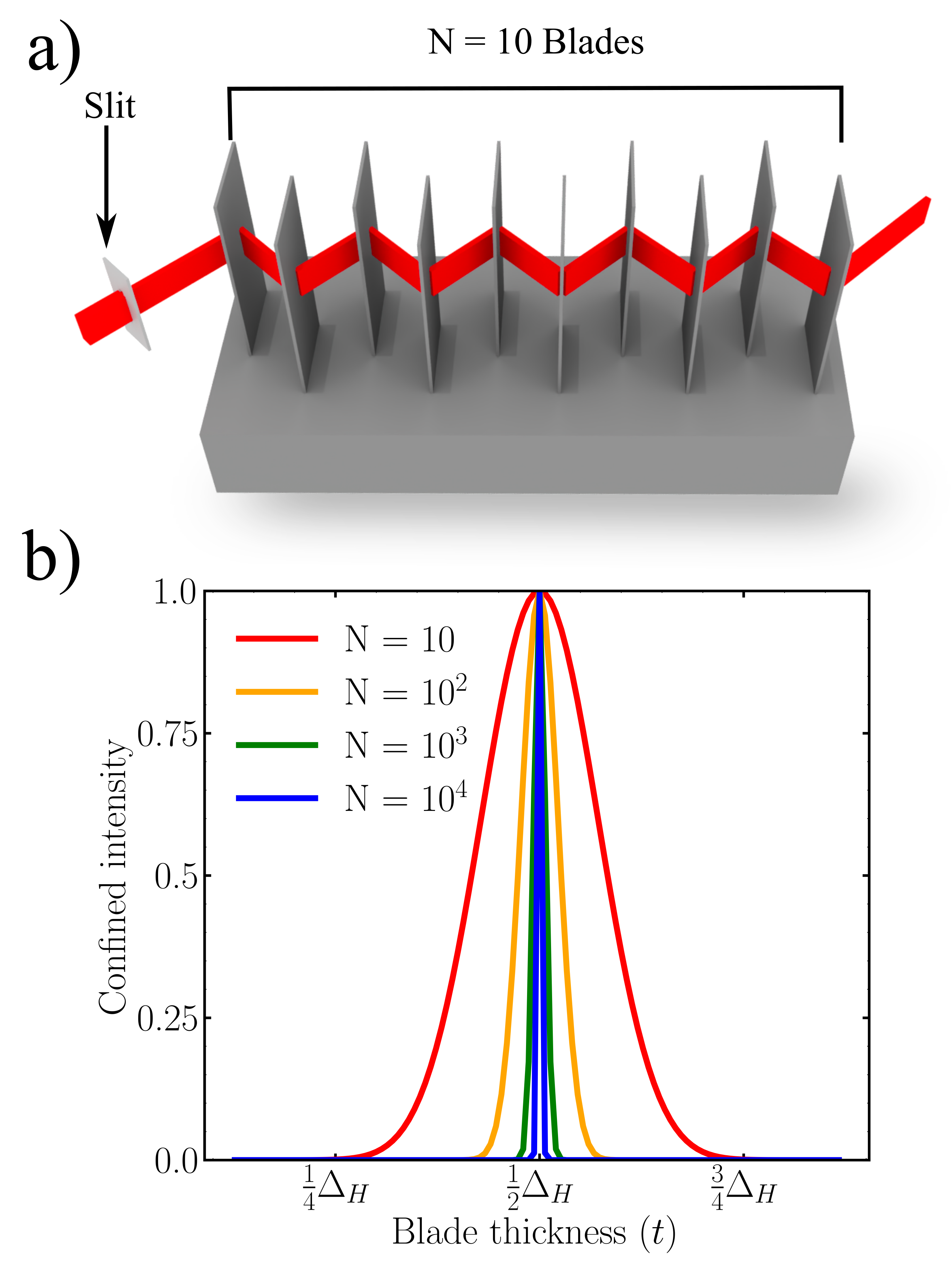}
    \caption{(a) Schematic of a Laue interferometer with $N = 10$ blades of identical thickness $t = \Delta_H/2$. (b) Simulation of the confined neutron intensity in the interferometer after $N$ reflections for varying blade thickness $t$. The blade thickness FWHM fabrication tolerances scale with blade number $N$ as $\Delta t_{FWHM} = \frac{1}{2}N^{-\frac{1}{2}} \Delta_H$.}
    \label{fig:fig6}
\end{figure}

Notably, it was proposed in Ref.~\cite{zeilinger1984symmetry} to develop interferometers composed of $N$ lamella (see Fig.~\ref{fig:fig6}a schematic), where each blade is of thickness $t = \Delta_H / 2$, as a method of producing measurable phase shifts from Schwinger scattering and the neutron electric dipole moment (nEDM). In this setup, each blade reflects a neutron plane-wave satisfying the Bragg condition with 100~\% probability. A uniform magnetic field is applied to introduce a spin flip between reflections, which ensures constructive addition of the spin rotation. For a measurement of the nEDM, its interaction with silicon must be isolated from Schwinger scattering by choosing an initial polarization and a spin flipping magnetic field oriented parallel to the spin-orbit magnetic field. The magnitude of the accumulated spin rotation for (220) reflection after $N$ blades is $\phi_{\text{spin}}(N)\approx 1.164\times10^{-2} \cdot N~\text{rad}$ for a spin-orbit measurement, and $\phi_{\text{spin}}(N)\approx 5.744 \times 10^{-11} \cdot N~\text{rad}$ for a nEDM measurement (assuming the current experimental limit of $d_n \sim 10^{-27} \cdot e ~ \text{cm}$~\cite{abel2020measurement}). For as few as $N=10$ blades, the spin-orbit interaction is magnified to a relatively large measurable phase shift of $\phi_{\text{spin}}(N)\approx0.1$~rad. The nEDM however, requires an unrealistic $N=10^6$ blades to even reach $\phi_{\text{spin}}(N)\approx10^{-4}$~rad.

To investigate the fabrication tolerances associated with such a unique interferometer geometry, we modeled the interferometer with a variable number of blades using the well-established quantum walk model for DD~\cite{Nsofini_2016, nsofini2017noise, nsofini2019coherence, nahman2022generalizing}. The validity of this approach for modeling neutron cavities has been demonstrated in Refs.~\cite{neutron_cav, lailey2026perfect}, while the implementation and treatment of plane-wave simulations are described in detail in Ref.~\cite{lailey2026unified}. Shown in Fig \ref{fig:fig6}b is the confined neutron intensity within the interferometer after $N$ Laue blades as the thickness of the blades are varied between zero and one~$\Delta_H$. As the number of blades increases, the tolerances on $t$ being exactly $\Delta_H/2$ become more strict. Extracting the FWHM of the simulated curves, the blade thickness fabrication tolerances scales with blade number $N$ as:
\begin{equation}
    \Delta t_{FWHM} = \frac{1}{2}N^{-\frac{1}{2}} \Delta_H.
\end{equation}
For $\Delta_H\sim50~\mu$m and $N=100$, confined neutron intensity is reduced by 1/2 when $t = \frac{\Delta_H}{2} \pm 1.25~\mu$m.

\section{Conclusion}
In summary, we have experimentally realized neutron and X-ray perfect crystal interferometers with blade thicknesses reaching the \pend length scale, demonstrating operation for devices with $t = 110~\mu$m and $t = 350~\mu$m. With X-rays, we observe fringe contrasts near $80~\%$ in single-shot measurements. These devices represent the thinnest neutron/X-ray interferometers realized to date and establish that multiple identical sub-millimeter lamella can be fabricated with sufficient uniformity to support high-contrast interference across centimeter-scale blade surfaces.

A central result of this work is the direct suppression of DD-induced beam spreading in all three blades. The $t = 110~\mu$m interferometer exhibits an approximately $8\times$ reduction in Borrmann fan broadening relative to the $t = 870~\mu$m device (at the same Bragg angle), enabling improved preservation of transverse phase information. This scaling highlights the advantage of operating in the pendell\"osung-length regime for neutron/X-ray phase-contrast imaging, where spatial resolution is otherwise limited by DD.

Operating at $t \sim \Delta_H$ enables access to a regime in which beam splitting is directly governed by the crystal thickness, providing a practical route to thickness-engineered interferometers with tailored splitting ratios. We extend our experimentally demonstrated three-blade interferometers to design an $N$-blade interferometer composed of identical blades with $t = \Delta_H/2$. In this configuration, we determine that fabrication tolerances scale as $\Delta t \propto N^{-1/2}$. While this scaling becomes increasingly stringent for large $N$, the required precision remains within the reach of current sub-micron fabrication techniques, supporting the feasibility of multi-blade architectures, including those proposed for enhanced neutron spin–orbit interactions. Quantitative estimates indicate that relatively modest blade numbers ($N \sim 10$) can already produce measurable spin-dependent phase shifts.

Notably, all measurements reported in this work were performed under non-isolated environmental conditions, with certain observed phase drift rates on the order of $\sim \pi / 0.1~\text{s}$. Despite this, high-contrast interference was observed indicating high intrinsic contrast of the devices. Future measurements at the NIST Neutron Interferometry and Optics Facilities, with full vibration and temperature stabilization, are expected to significantly suppress phase noise and recover near-unity contrast, enabling a more precise characterization of performance limits in this regime.
 
These results enable new opportunities in high-resolution phase imaging, customized beam-splitting control, exploration of different interferometer geometries, and precision measurements, including interferometric implementations of spin-orbit coupling and related fundamental symmetry tests.

\section*{Acknowledgments}
This work was supported by the Canadian Excellence Research Chairs (CERC) program, the Natural Sciences and Engineering Research Council of Canada (NSERC) Discovery program, the NSERC Canada Graduate Scholarships programs (CGS-M and PGS-D), the Canada  First  Research  Excellence  Fund  (CFREF), and the US Department of Energy, Office of Nuclear Physics, under Interagency Agreement 89243019SSC000025.  We acknowledge the Diamond Light Source for time on B16 under proposal OM43923-1. The authors would like to thank B16 beamline technician Andrew Malandain for their support during the experiment. Experiments at the ISIS Neutron and Muon Source were supported by beamtime allocation RB2510504 from the Science and Technology Facilities Council. The authors would like to thank Ritik Kotak for building the temperature monitoring setup used at ISIS. A portion of this research used resources at the High Flux Isotope Reactor, a DOE Office of Science User Facility operated by the Oak Ridge National Laboratory. The neutron beam time was allocated to CTAX under proposal number IPTS-34324.1. Work at the University of Toronto was supported by the Canada Foundation for Innovation (CFI) and the Government of Ontario for Project No. 36404.

\bibliographystyle{ieeetr}
\bibliography{mybib.bib}

\newpage
\section*{Appendix}

\begin{figure}[ht]
    \centering
    \includegraphics[width=1\linewidth]{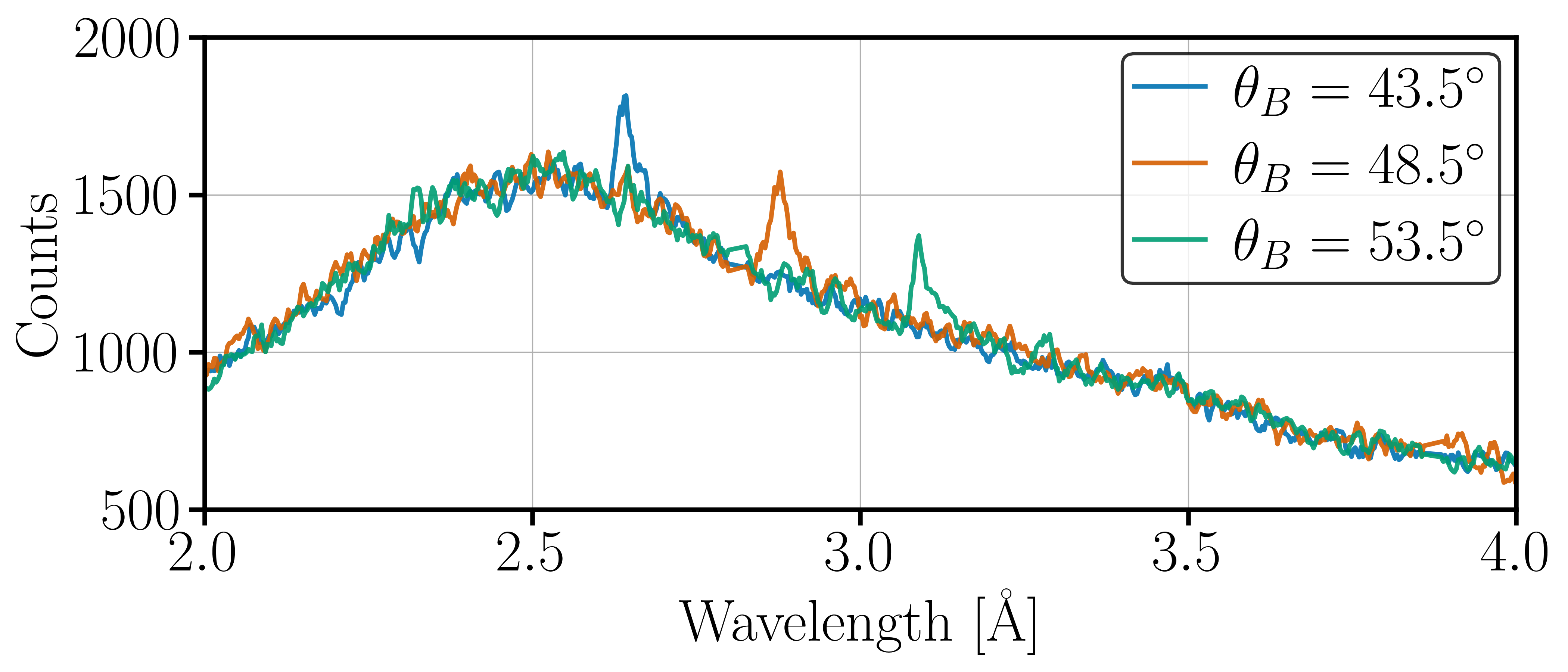}
    \caption{O-beam counts across the wavelength spectrum for \PCItwo~for several Bragg angles. The TOF modality enables selection of different effective wavelengths by rotating only the interferometer. Counts are integrated over a roughly $12~\text{mm} \times 2~\text{mm}$ region. The resulting wavelength profiles are smoothed using a rolling average over 5 bins, corresponding to approximately $0.01~\text{\AA}$.}
    \label{fig:fig7}
\end{figure}

Time-of-flight (TOF) measurements at the ISIS IMAT beamline enabled selection of different wavelengths from the polychromatic spectrum by rotating the interferometer (see Fig.~\ref{fig:fig7}). Rotation of the interferometer about the center of the entrance blade allowed access to Bragg angles of $43.5^\circ$, $48.5^\circ$, and $53.5^\circ$, corresponding to distinct peaks in the wavelength-resolved intensity.

\begin{figure}[ht]
    \centering
    \includegraphics[width=1\linewidth]{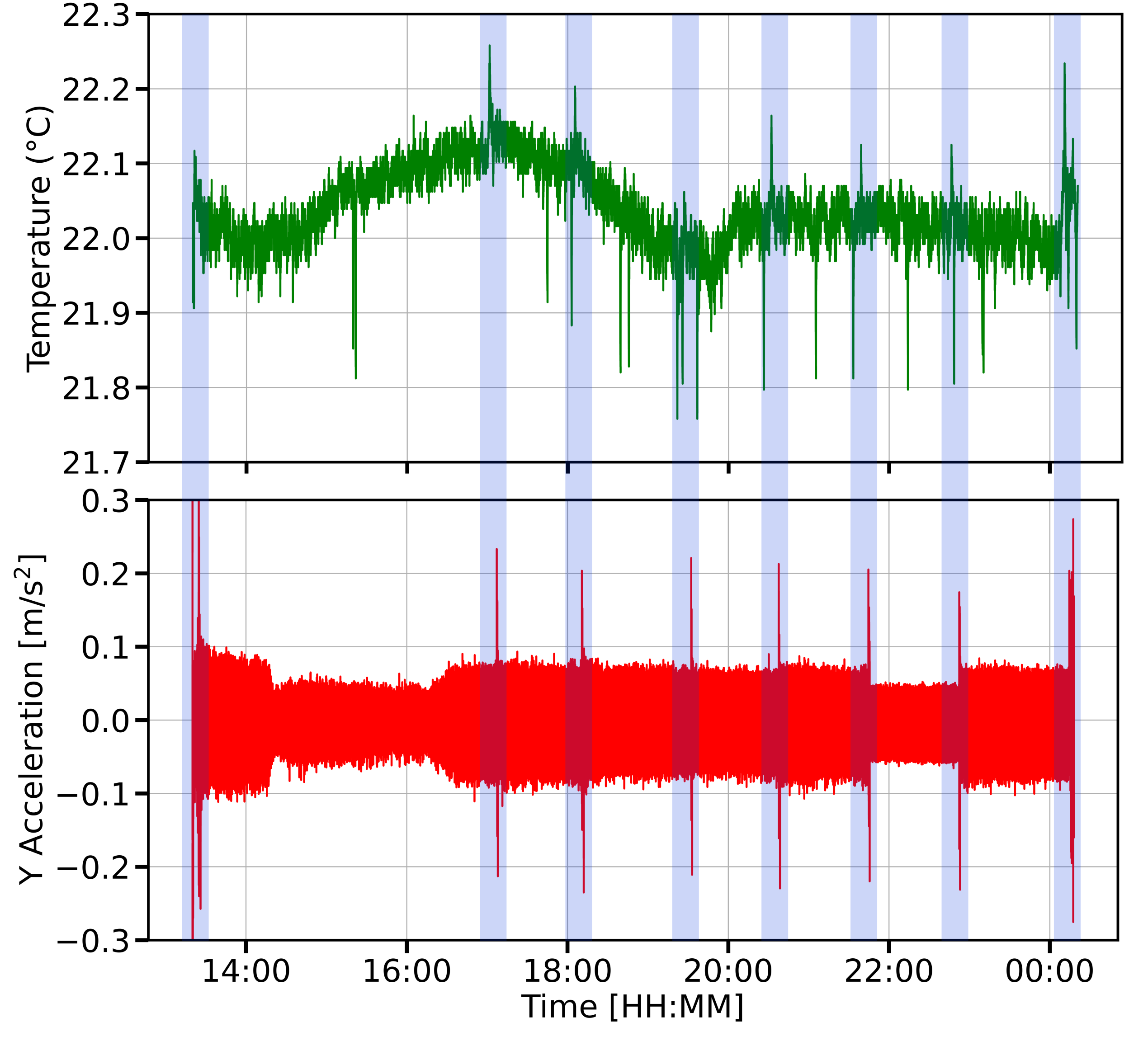}
    \caption{Temperature and vibration monitoring during the contrast measurement for \PCIthr~over approximately 11~h. Temperature spikes of $\sim 0.1^\circ$ and vibration spikes up to $\sim 0.2~\text{m/s}^2$ occur each time the hutch is entered to rotate the phase flag (blue shaded regions). The average temperature was $22.04^\circ \pm 0.05^\circ$. Fourier analysis of the accelerometer data shows a dominant frequency at 50~Hz, consistent with the mains electrical frequency. The temperature sensor and accelerometer were placed adjacent to the interferometer.}
    \label{fig:fig8}
\end{figure}

For the contrast measurement at IMAT (Fig.~\ref{fig:fig5}c), the phase flag rotation was performed manually. This introduced significant variations in temperature and vibration, as shown in Fig.~\ref{fig:fig8}. The experimental hutch was entered approximately once per hour over a $\sim 11$~h measurement to rotate the phase flag. As indicated by the shaded blue regions in Fig.~\ref{fig:fig8}, each intervention produced temperature excursions of approximately $0.1^\circ$ and vibration spikes up to $\sim 0.2~\text{m/s}^2$. These disturbances contribute to phase drift in the interferometer and reduce the measured contrast. Under controlled environmental conditions, the same interferometer previously achieved a contrast of $92~\%$, compared to $58~\%$ obtained here under non-isolated conditions.

\clearpage

\end{document}